\documentclass[reprint, amsmath,amssymb, aip ,twocolumn, superscriptaddress,longbibliography]{revtex4-2}
\usepackage[utf8]{inputenc}
\usepackage[english]{babel}
\usepackage{hyperref}
\usepackage{dcolumn}
\usepackage{bm}
\usepackage{hyperref}

\usepackage[dvipsnames]{xcolor}
\usepackage{graphicx}
\usepackage[capitalise]{cleveref}
\usepackage{textcomp}
\usepackage{lipsum}
\usepackage{pdfcomment}
\usepackage{bold-extra}
\usepackage{ulem}

\newcommand{\greg}[1]{#1}

\newcommand{\ks}[1]{\textcolor{OliveGreen}{#1}}

\newcommand{\beginsupplement}{%
 \setcounter{table}{0}
 \renewcommand{\thetable}{S\arabic{table}}%
 \setcounter{figure}{0}
 \renewcommand{\thefigure}{S\arabic{figure}}%
 \renewcommand{\thesubsection}{S-\arabic{subsection}}
}
\usepackage{titlesec}
\titleformat*{\section}{\bfseries\Large}
\titleformat*{\subsection}{\bfseries}

\usepackage{caption}
\begin{document}

%  _____ ___ _____ _     _____
% |_   _|_ _|_   _| |   | ____|
%   | |  | |  | | | |   |  _|
%   | |  | |  | | | |___| |___
%   |_| |___| |_| |_____|_____|
%

\title{Tailoring Broadband Kerr Soliton Microcombs via Post-Fabrication Tuning of the Geometric Dispersion}
% \title{Tailoring Dissipative Kerr Soliton Microcombs Through Fine Thickness Trimming}
% \title{Precised Thickness Post Fabrication Trimming Of Mircoresonator Thickness for Microcomb Tailoring}

%     _   _   _ _____ _   _  ___  ____  ____
%    / \ | | | |_   _| | | |/ _ \|  _ \/ ___|
%   / _ \| | | | | | | |_| | | | | |_) \___ \
%  / ___ \ |_| | | | |  _  | |_| |  _ < ___) |
% /_/   \_\___/  |_| |_| |_|\___/|_| \_\____/
%
\author{Gregory Moille}
\email{gmoille@umd.edu}
\affiliation{Joint Quantum Institute, NIST/University of Maryland, College Park, MD 20742, USA}
\affiliation{Microsystems and Nanotechnology Division, National Institute of Standards and Technology, Gaithersburg, MD 20899, USA}
\author{Daron Westly}
\affiliation{Microsystems and Nanotechnology Division, National Institute of Standards and Technology, Gaithersburg, MD 20899, USA}
\author{Ndubuisi George Orji}
\affiliation{Microsystems and Nanotechnology Division, National Institute of Standards and Technology, Gaithersburg, MD 20899, USA}
\author{Kartik Srinivasan}
\affiliation{Joint Quantum Institute, NIST/University of Maryland, College Park, MD 20742, USA}
\affiliation{Microsystems and Nanotechnology Division, National Institute of Standards and Technology, Gaithersburg, MD 20899, USA}
\date{\today}

%     _    ____ ____ _____ ____      _    ____ _____
%    / \  | __ ) ___|_   _|  _ \    / \  / ___|_   _|
%   / _ \ |  _ \___ \ | | | |_) |  / _ \| |     | |
%  / ___ \| |_) |__) || | |  _ <  / ___ \ |___  | |
% /_/   \_\____/____/ |_| |_| \_\/_/   \_\____| |_|
%
\begin{abstract}
\noindent Geometric dispersion in integrated microresonators plays a major role in nonlinear optics applications, especially at short wavelengths, to compensate the natural material normal dispersion. Tailoring of geometric confinement allows for anomalous dispersion, which in particular enables the formation of microcombs which can be tuned into the dissipative Kerr soliton (DKS) regime. Due to processes like soliton-induced dispersive wave generation, broadband DKS combs are particularly sensitive to higher-order dispersion, which in turn is sensitive to the ring dimensions at the nanometer-level. For microrings exhibiting a rectangular cross section, the ring width and thickness are the two main control parameters to achieve the targeted dispersion. The former can be easily varied through parameter variation within the lithography mask, yet the latter is defined by the film thickness during growth of the starting material stack, and can show a significant variation (few percent of the total thickness) over a single wafer. In this letter, we demonstrate that controlled dry-etching allows for fine tuning of the device layer (silicon nitride) thickness at the wafer level, allowing multi-project wafers targeting different wavelength bands, and post-fabrication trimming in air-clad ring devices. We demonstrate that such dry etching does not significantly affect either the silicon nitride surface roughness or the optical quality of the devices, thereby enabling fine tuning of the dispersion and the spectral shape of the resulting DKS states. 
\end{abstract}
\maketitle

%  ___ _   _ _____ ____   ___
% |_ _| \ | |_   _|  _ \ / _ \
%  | ||  \| | | | | |_) | | | |
%  | || |\  | | | |  _ <| |_| |
% |___|_| \_| |_| |_| \_\\___/
%

Integrated microresonators have been demonstrated in a variety of different materials in which the frequency spacing of a set of optical modes, the free spectral range (FSR), increases with the optical frequency, a behavior termed as anomalous dispersion~\cite{kippenberg_dissipative_2018, ilchenko_dispcomp}. This allows for compensation of the intensity-dependent Kerr frequency shift, enabling the fundamental conservation of both energy and momentum - or frequency and angular momentum in the special case of periodic whispering gallery modes structures. This fundamental dispersion property allows for a variety of $\chi^{(3)}$ nonlinear processes to be effectively realized~\cite{ferrera_fwm_ring,liang_comb_wgm}. In particular, four-wave mixing (FWM) and optical parametric oscillation from a continuous-wave pump can occur, seeding the so-called modulation instability frequency microcomb~\cite{savchenkov_comb_caf2}, which is a first step in obtaining a stable pulse state in the dissipative Kerr soliton (DKS) regime~\cite{chembo_modal_expansion}. In photonic platforms where DKS states have been demonstrated, for example in microrings~\cite{suh_ghzreprate_sio2, brasch_dsk_cherenkov, gong_highfidelity_aln, he_selfstarting_linbo3, moille_DKS_III-V}, material dispersion (which is normal) is compensated by geometric dispersion, i.e. the confinement of the guided mode within the structure relative to the wavelength, to obtain anomalous dispersion~\cite{okawachi_bandwidth_2014}. The amount of dispersion compensation needed is particularly acute at shorter wavelengths as photonic materials, such as Si$_3$N$_4$, SiO$_2$, AlN or LiNbO$_3$, exhibit pronounced normal dispersion in the visible range. %

However, obtaining anomalous dispersion is usually not sufficient for supporting a DKS state that suits application needs. In particular, for full comb stabilization requiring an octave of bandwidth, the DKS spectrum has to be extended using dispersive waves (DWs)~\cite{spencer_optical-frequency_2018, yu_tuning_2019,gong_nearoctave_linbo3}, a phenomenon relying on higher-order dispersion~\cite{brasch_dsk_cherenkov, cherenkov_dissipative_2017}. In this case, a DKS comb tooth overlaps with a cavity resonance far away from the pumped mode, exhibiting resonant enhancement at this particular wavelength, and increasing the comb power locally. The spectral positions of such DWs are highly sensitive to the geometric dispersion, which in turn is controlled primarily through the ring width~\cite{yu_tuning_2019} and the ring thickness~\cite{black_disp_tantala}. Dimensional control down to the few nanometer level is required for precise targeting of DW spectral locations. The ring width, being an in-plane parameter, can be addressed at the design level where different structures on the same chip are given a slight variation in their ring width. However, thickness is defined during growth and requires subsequent post-growth processes if it is to be adjusted.

\begin{figure}[!t]
	\includegraphics[width=\columnwidth]{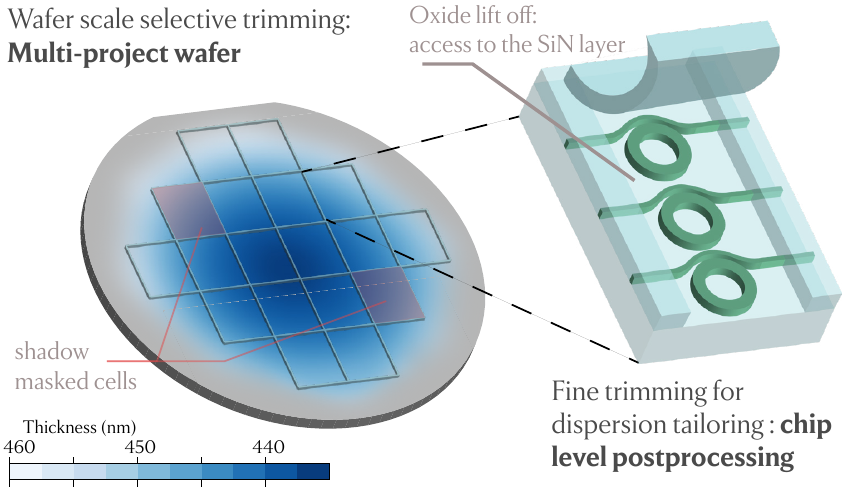}
	\caption{\label{fig:1} A 100~mm wafer of Si\textsubscript{3}N\textsubscript{4}/SiO\textsubscript{2} on a Si substrate that is divided into several cells, each containing a certain number of die in a multi-project fabrication run. The heat map corresponds to the Si$_3$N$_4$ thickness measured by a spectroscopic ellipsometer and exhibiting a variation of more than $\approx$~30~nm ($\pm$3.5~\% of the mean thickness of 445~nm) across the whole wafer. Cells can be selectively masked for dry-etch trimming of particular areas of the wafer, allowing for different controlled thicknesses across the wafer to accommodate different projects. For devices that lack a top-cladding material (air-clad structures), fine trimming can also occur after the completed device fabrication, with measurement feedback guiding the level of trimming needed. In our case, we lift-off the SiO\textsubscript{2} (right image) to uncover the rings for dispersion and fine trimming purposes while keeping SiO$_2$ at the facets for improved fiber input/output coupling.}
\end{figure}%

\begin{figure}[!t]
    \includegraphics[width=\columnwidth]{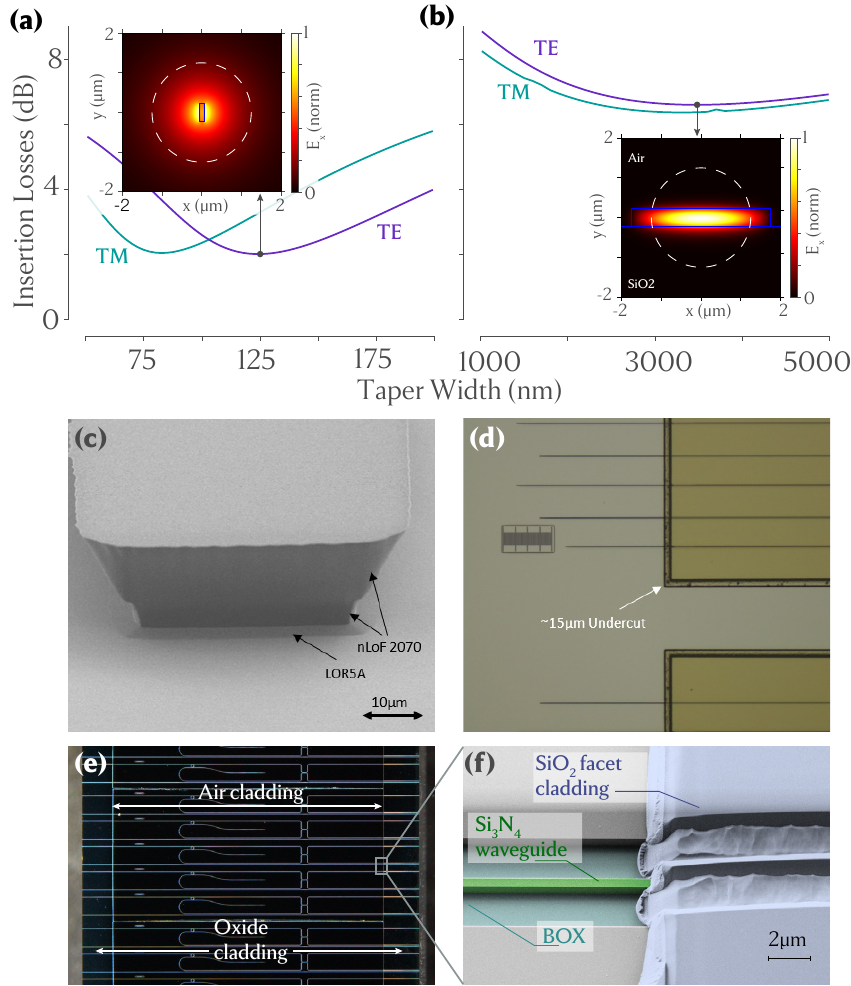}
    \caption{\label{fig:2} \textbf{(a)-(b)} Insertion loss calculation computed through finite element simulations, assuming a spot-size diameter of the lensed fiber of 2.5~$\mu$m and assuming 430~nm thick Si$_3$N$_4$, for (a) a fully SiO$_2$-clad waveguide and (b) an air top-cladding waveguide. For each structure, TE (\greg{purple}) and TM (teal) insertion losses are calculated and the mode profile for the optimized dimension for TE \greg{polarization} are showcased, along with the structure (black) and the lensed fiber spot size (dashed white). The measured insertion losses for SiO$_2$ clad waveguides is between 3.5~dB to 4.0~dB per facet.\textbf{(c)} Scanning electron microscope (SEM) image and \textbf{(d)} optical microscope image of the dual-layer resist used to lift-off the low-temperature PECVD SiO$_2$ to create selectively-clad photonic chips. \textbf{(e)} Optical microscope and \textbf{(f)} SEM images after the SiO$_2$ lift-off process, which uncovers the ring area while keeping the SiO$_2$ at the chip facets for low insertion losses.  The SEM image in (f) highlights smooth transition between the air and oxide clad regions. }
\end{figure}

%In particular with material under stress, such as Si\textsubscript{3}N\textsubscript{4} where deposition is usually non-uniform over a full wafer and exhibit a non-neglible variation of thickness depending where the chip is fabricated (see ~\cref{fig:1}). Therefore, large dispersion change can occur for two chips fabricated on the same exact run, yet different position on the wafer. \\

%Such adjustments can be considered in a couple of different contexts, for example, to overcome the typical non-uniformity in low pressure chemical vapor deposition of Si$_3$N$_4$ (usually at the percent-level or higher~\cite{gumpher_lpcvd_2003}), or to accommodate... 

In this letter, we demonstrate a dry etch trimming approach that enables fine tuning of the thickness of Si$_3$N$_4$ microresonator frequency comb devices with a resolution down to the few nanometer level (\cref{fig:1}). One application of this trimming is in chip-to-chip compensation of the natural thickness variation across a wafer, which is typically at the percent level or higher for growth via low pressure chemical vapor deposition~\cite{gumpher_lpcvd_2003}. The dry etch trimming can also be used to realize thickness differences of several tens of nanometers, enabling multi-project wafers in which different thicknesses are needed to target, for example, frequency combs operating in different spectral bands or other $\chi^{(3)}$ nonlinear nanophotonic devices. Working in an air-clad system in which the Si$_3$N$_4$ device layer remains accessible after fabrication (see~\cref{fig:1}), we perform optical characterization of the microcomb devices before and after trimming steps.  We show that the frequency shift of the resonator modes due to dry etch trimming can be repeatably controlled and does not adversely impact their optical quality factor ($Q$). We further show how DKS microcomb states, and in particular, DW positions, are modified through thickness trimming, with the DW shift predictable based on the impact of device layer thickness on higher-order dispersion.

%   ___       _     _         ____ _           _ 
%  / _ \__  _(_) __| | ___   / ___| | __ _  __| |
% | | | \ \/ / |/ _` |/ _ \ | |   | |/ _` |/ _` |
% | |_| |>  <| | (_| |  __/ | |___| | (_| | (_| |
%  \___//_/\_\_|\__,_|\___|  \____|_|\__,_|\__,_|

%Fine trimming of the thickness at the wafer level is still relatively easy as the full chip is not yet fabricated and this step can happen at any time. Shadow masking the part of the wafer corresponding to the different thickness region of interest, for instance allowing two different to be made on the same fabrication run. Post-fabrication of individual chips are however more compelling. 

Most of the reported DKS microcombs in Si\textsubscript{3}N\textsubscript{4} resonators have been demonstrated in devices that are fully clad with SiO$_2$~\cite{brasch_dsk_cherenkov,yu_tuning_2019}. This approach has the advantages of making the resonator relatively insensitive to its environment and allows for better heat dissipation. However, post-fabrication steps to modify the geometry based on measurement results are made impossible as the resonator is not directly accessible. Here, we use a system where the ring resonator does not have any material as the top cladding. Such air-clad resonators have been demonstrated to support high-$Q$ and octave-spanning microcombs~\cite{li_stably_2017, moille_cryogenic}, as well as other wide-band nonlinear processes~\cite{li_stably_2017,lu_-chip_2020}. Yet in our system, the facet waveguides are still embedded in SiO$_2$ (\cref{fig:2}). This selectively-clad chip thus enables direct access to the microresonator section for trimming purposes, while providing more optimized facet coupling regions for improved insertion losses from/to lensed optical fibers (\cref{fig:2}(a)). In particular, the full SiO$_2$ cladding both provides for a more symmetric mode than what would be possible with an air-clad waveguide and ensures that a guided mode is supported even when inverse tapers are used to expand the mode size to best match that of lensed optical fibers (asymmetrically clad waveguides are not guaranteed to support even a fundamental guided mode if the taper width is too small). Together, this results in insertion losses that are predicted to be as low as 2~dB per facet at 1060~nm (Fig.~\ref{fig:2}(a)), an improvement on the 6~dB per facet loss that could be reached in air-clad waveguides tapered to a larger width at the chip facets (Fig.~\ref{fig:2}(b)). In practice, we typically measure insertion losses between 3.5~dB to 4~dB per facet for a nominal inverse taper width of 150~nm (see supplementary material). We note that the Fresnel reflection between the two cladding regions, which could potentially lead to other insertion losses, is negligible and in the 0.1~dB range (see supplementary material). The selective SiO$_2$ cladding is realized through a photoresist lift-off process (Fig.~\ref{fig:2}(c)-(d)), where removal of the top SiO\textsubscript{2} cladding over the microring is made possible through low-temperature (180$^{\circ}$C) inductively-coupled plasma enhanced chemical vapor deposition (PECVD) of SiO$_2$, allowing photoresist to be used without polymerization that would occur at high temperature. We use a dual layer of LOR5-A and nLOF resists~\cite{NIST_disclaimer} (one spin coating of LOR5-A, two spin coatings of nLOF), thicker than the targeted deposited SiO$_2$ thickness of 3~{\textmu}m, and optically expose them over the microring regions of the chip. Optical and scanning electron microscope images of a completed chip are shown in Fig.~\ref{fig:2}(e)-(f).

\begin{figure}[t]
    \includegraphics[width=\columnwidth]{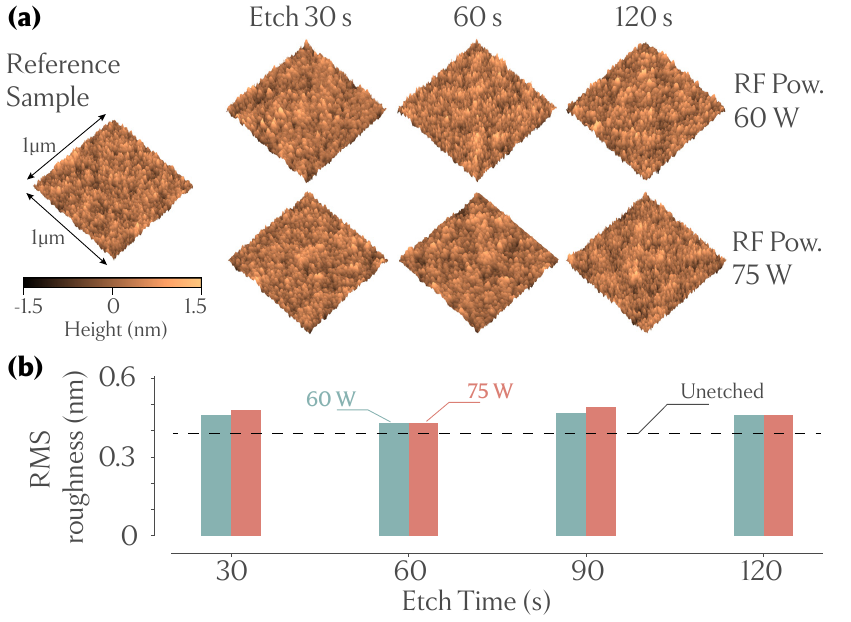}
    \caption{\label{fig:3} \textbf{(a)} Atomic force microscopy measurement across a 1~$\mu$m\textsuperscript{2} area of the top surface of a Si$_3$N$_4$ film before and after RIE trimming steps, the latter of which vary in duration and radio frequency (RF) power. \textbf{b} Root-mean square (RMS) roughness comparison for different etch duration (horizontal scale) and a RF power of 60~W (green) and 75~W (red). \greg{The roughness standard deviation for each sample is between 38~pm and 45~pm and is too small to be displayed here.} The black dashed line indicates the RMS roughness prior to trimming.}
\end{figure}

%  _____     _                     _             
% |_   _| __(_)_ __ ___  _ __ ___ (_)_ __   __ _ 
%   | || '__| | '_ ` _ \| '_ ` _ \| | '_ \ / _` |
%   | || |  | | | | | | | | | | | | | | | | (_| |
%   |_||_|  |_|_| |_| |_|_| |_| |_|_|_| |_|\__, |
%                                          |___/ 
As the above method leaves selected regions of the Si$_3$N$_4$ layer accessible, a reactive ion etch (RIE) can be used to trim the Si\textsubscript{3}N\textsubscript{4} thickness in those regions. In comparison to wet etching of Si$_3$N$_4$, which can be done through dilute hydrofluoric acid or heated phosphoric acid but are isotropic (and hence impact both resonator in-plane dimensions and thickness), we emphasize that dry etching can be tailored to predominantly impact thickness. In particular, we use a CF\textsubscript{4} and O\textsubscript{2} plasma mixture at \greg{2~Pa (15~mTorr)} chamber pressure, with \greg{$5\times10^{7}$~m\textsuperscript{3}.s\textsuperscript{-1} (30~sccm) and $3.33\times10^{7}$~m\textsuperscript{3}.s\textsuperscript{-1}  (20~sccm)} flow respectively, at room temperature. In comparison to inductively-coupled plasma dry etching with a significant chemical component, this RIE etch is mostly physical and only affects the top surface and not the sidewalls of the microrings. To ensure that the RIE does not adversely impact microring top surface roughness, we perform atomic force microscope (AFM) measurements of the Si$_3$N$_4$ surface for several films that have been etched for times ranging between 30~s to 120~s and for two different radio frequency (RF) powers (Fig.~\ref{fig:3}). The AFM measurements were made in tapping mode using wear-resistant high density diamond-like carbon tips of 2~nm radius. To further minimize tip wear and ensure imaging consistency, each surface map was acquired using a fresh tip. These measurements across a region of a reference (unetched) Si$_3$N$_4$ sample are shown in Fig.~\ref{fig:3}(a), where the root-mean square (RMS) of the surface roughness is 0.39~nm. The etched samples show a slight increase in RMS roughness to $\approx$~0.46~nm, with approximately no increase in rms roughness for longer etch times or the higher RF power \greg{(Fig.~\ref{fig:3}(b))}. Importantly, this rms roughness remains significantly below the sidewall roughness measured for high-$Q$ Si$_3$N$_4$ devices~\cite{robert_sidewallroughness}, and is comparable to the Si$_3$N$_4$ roughness after film growth \greg{(and before chemical-mechanical polishing) reported by Ji \textit{et al.}  (0.38~nm)}~\cite{ji_lowthreshold}.

%It is interesting to note that this roughness value is order of magnitude lower than sidewall roughness, which is usually close to 2~nm~\cite{robert_sidewallroughness}, and actually remains better the roughness of freshly grown nitride reported by Ji \textit{et al.}~\cite{ji_lowthreshold} before chemical-mechanical polishing (0.08~nm). 

\begin{figure*}[t]
    \includegraphics[width=\linewidth]{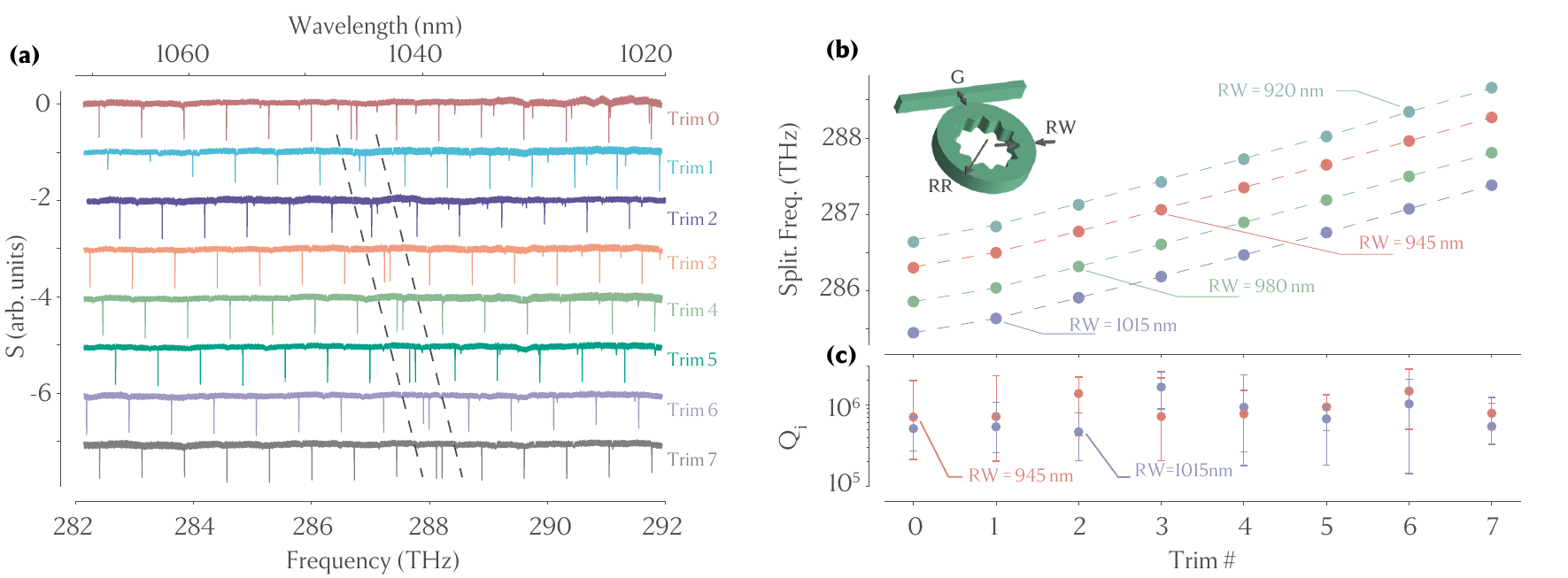}
    % Linear transmition spectra of a ring resonator for $RW=920$~nm with an initial thickness of $H=430$~nm, undergoing 7 identical trimming process. The mode splitted resonance that are designed at $M=$ allows to track the frequency shift with the variation of thickness. \textbf{(b)} Center frequency of the splitted mode with RW (colors) and triming steps \textbf{(c)} Intrinsic quality average ofer all the resonance in the linear scan badnwidth (dot) for $RW=945$~nm$RW=1015$~nm (blue). Scale bar represent the variance across all the resonances in quality factors.}
    \caption{\label{fig:4} \textbf{(a)} Linear transmission data for a microring resonator ($RW=920$~nm) with an initial thickness of $H=430$~nm, after several different trimming steps, where for each step the trim was for 20~s and with 45~W of RF power. The frequency-split mode indicated in-between the dashed black lines is the result of an inner sidewall modulation whose period was set to target a specific mode of the ring (\greg{azimuthal mode number} $M=237$), providing an easy reference by which mode frequencies can be tracked. (b) Frequencies of the tracked microring mode for several different $RW$ values and several different trim steps, showing a nearly constant frequency shift per step. (c) Average intrinsic quality factor ($Q_\text{i}$) of 14 different resonances at each trim step and for $RW=945$~nm and $RW=1015$~nm. The error bars represent one standard deviation values based on the spread in the $Q_\text{i}$ values.}
\end{figure*}

%   ___        _   _           _   ____                   
%  / _ \ _ __ | |_(_) ___ __ _| | |  _ \ _ __ ___  _ __   
% | | | | '_ \| __| |/ __/ _` | | | |_) | '__/ _ \| '_ \  
% | |_| | |_) | |_| | (_| (_| | | |  __/| | | (_) | |_) | 
%  \___/| .__/ \__|_|\___\__,_|_| |_|   |_|  \___/| .__(_)
%       |_|                                       |_| 
\begin{figure*}[t]
    \includegraphics[width=\textwidth]{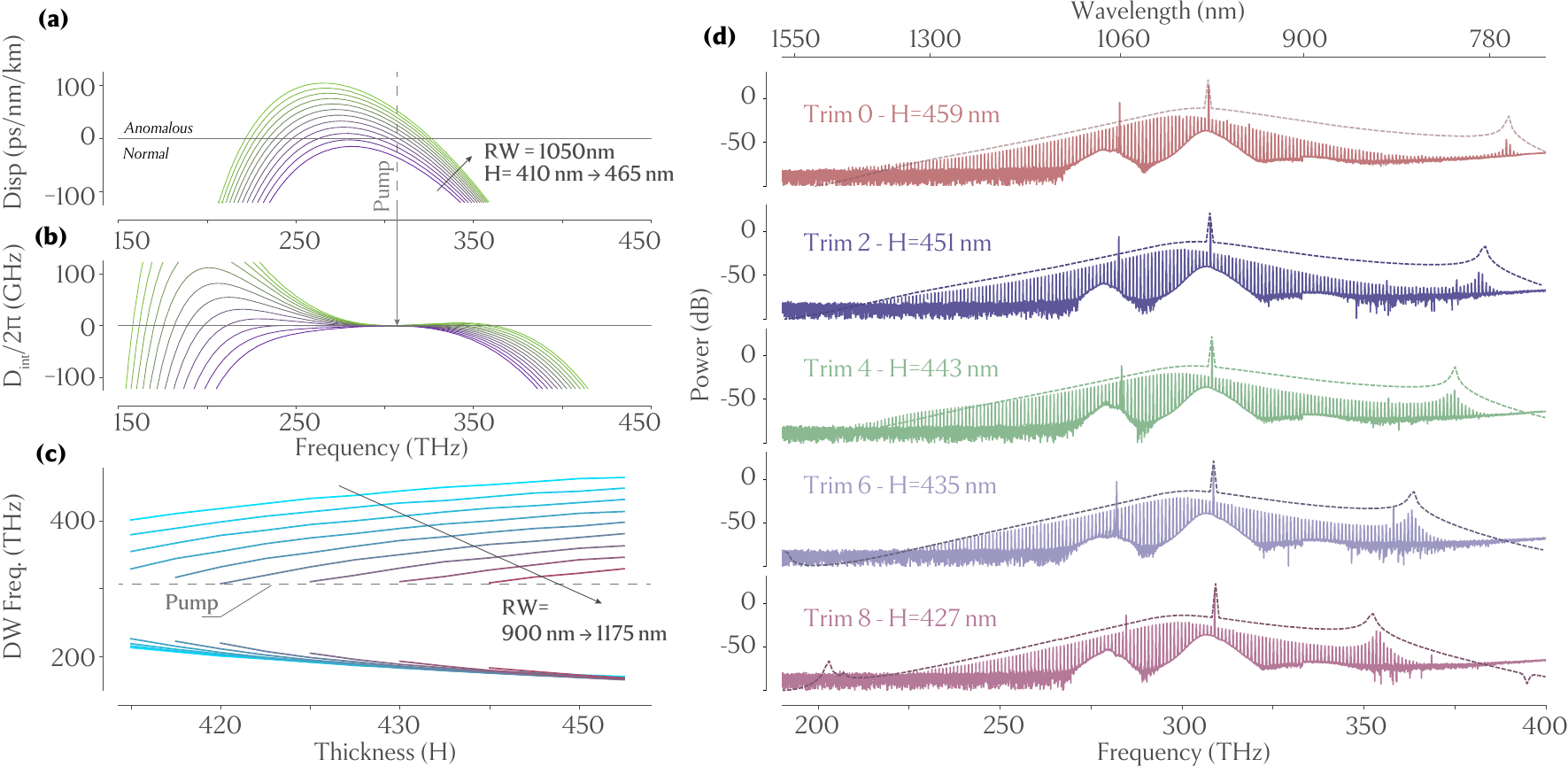}
    \caption{
    \label{fig:5}
    \textbf{(a)} Dispersion parameter of the Si$_3$N$_4$ microring resonator for $RR=23$~{\textmu}m, $RW=1050$~nm, and for different thicknesses between $H$=410~nm and $H$=465~nm,  highlighting the crossover from normal to anomalous dispersion for sufficiently thick films. \textbf{(b)} Integrated dispersion of the same ring resonator at 305~THz (983~nm) pump frequency. \textbf{(c)} Position of the DWs from the integrated dispersion zero crossing for a 305~THz pump frequency. \textbf{(d)} DKS microcomb state measured by driving the same pumped mode at each trimming step of -4~nm in thickness (every other DKS spectrum is displayed, i.e., after -8~nm steps). The DW at high frequency (short wavelength) is tuned at a rate of $\approx$~-73~GHz/nm. The dashed lines represent the obtained spectra through LLE simulations, where the only change in input for each spectrum is the shift in dispersion corresponding to a change of -4~nm thickness per trimming step. The theoretical and experimental DWs agree closely. The power discrepancy is due to the chromatic coupling of the straight waveguide coupler, resulting in undercoupling at  short wavelengths. Soliton stabilization is provided by a counterpropagating 283~THz (1060~nm) pump.}
  \end{figure*}
  
While the AFM suggests that RIE trimming results in at most a small increase in the Si$_3$N$_4$ film roughness, it is still important to understand the impact of the trimming on the optical properties of the resonator modes, in particular, the influence on resonator $Q$ and modal frequencies, and the extent to which the trimming can be repeatably applied. To do so, we need to track the same resonator modes before and after trimming, which can be difficult in microring resonators due to the large number of modes present.  We use the selective mode splitting technique employed in Ref.~\onlinecite{lu_mms}, where the ring resonator has an inner sidewall modulation with a period set to couple and frequency split the initially degenerate clockwise and counterclockwise traveling wave modes with a specific azimuthal mode number.  Here, we target the azimuthal mode number of 237 ($2\times237$ periods along the ring inner circumference), resulting in a resonance frequency that is split close to 287.5~THz (Fig.~\ref{fig:4}(a)). Using the \greg{trim etch recipe} presented earlier, we perform a series of seven trims for devices which have been designed with different nominal ring width ($RW$) varying from 920~nm to 1015~nm~on the same chip, with each trim step lasting 20~s and at 45~W of RF power. \greg{The reduced RF power relative to that used in the AFM measurements in Fig.~\ref{fig:3} provided finer trimming resolution, while ensuring that the induced surface roughness was minimized.} We observe a nearly uniform frequency shift of $\approx$~0.29~THz with each step for each of the considered $RW$ values (Fig.~\ref{fig:4}(b)), though there are some small discrepancies from this trend, for example, the first trim step produced a somewhat smaller shift than the subsequent ones. Using finite element method eigenmode simulations, we retrieve the thickness dependence of the resonance frequencies ${\partial}f_\mathrm{res}/{\partial}H=-73.2$~GHz/nm for the $RW$ range under study (from 900~nm to 1100~nm), and from the frequency shifts measured in Fig.~\ref{fig:4}(b), we estimate that each 20~s trim step reduces the thickness by 4~nm. Even finer trimming steps (e.g., 2~nm for a 10~s etch) should be possible; however, we note that a linear scaling of etched amount with etch time should not necessarily be assumed due to plasma ignition delay (for short etch times) and sample heating (for long etch times). Finally, we consider the intrinsic quality factor ($Q_\text{i}$) for 14 resonances in our linear frequency scan for each trimming step. We find that the average $Q_\mathrm{i}$ remains about the same (within the one standard deviation uncertainty values determined by the spread in the $Q_\text{i}$ data). This is consistent with the AFM measurements indicating that the top surface roughness is only slightly increased by the trimming, and suggests that the post-fabrication trimming process can provide a method to tune the geometric dispersion of devices without significantly impacting their losses, which is critical for nonlinear applications.

%  ____  _                         _             
% |  _ \(_)___ _ __   ___ _ __ ___(_) ___  _ __  
% | | | | / __| '_ \ / _ \ '__/ __| |/ _ \| '_ \ 
% | |_| | \__ \ |_) |  __/ |  \__ \ | (_) | | | |
% |____/|_|___/ .__/ \___|_|  |___/_|\___/|_| |_|
%             |_|                                

Fine tuning of the thickness of a ring resonator not only allows for fine shifting of the frequency of resonances - which can be of interest for post-fabrication tuning to overlap with atomic transitions or other frequencies of interest - but also impacts the overall dispersion of the resonator, in particular the higher order dispersion coefficients responsible for dispersive wave creation~\cite{brasch_dsk_cherenkov}. For such microcombs, it is convenient to introduce the integrated dispersion $D_\mathrm{int}(\mu) = \omega_\mathrm{res}(\mu) - \omega_\mathrm{DKS}$, where $\mu$ is the mode number relative to the pumped mode (\textit{i.e.} $\mu_\mathrm{pmp} = 0$), which effectively measures the discrepancy between the resonator mode frequencies and the supported DKS frequency comb teeth $\omega_\mathrm{DKS} \approx \omega_{pmp} + D_1 \mu$, which are spaced by the resonator free spectral range that is approximately given by $D_{1}/2\pi$. As the modal confinement is partly set by the resonator thickness, the resonator dispersion (calculated in Fig.~\ref{fig:5}(a)) and therefore its integrated dispersion (calculated in Fig.~\ref{fig:5}(b)) can be significantly impacted by modifying its thickness for a fixed ring width. This ultimately results in a variation in the frequency of the DWs (Fig.~\ref{fig:5}(c)) on both the low and high frequency sides of the spectrum. DWs are crucial elements in realizing octave spanning DKS frequency combs that can be self-referenced~\cite{spencer_optical-frequency_2018} and reach atomic transition frequencies~\cite{yu_tuning_2019}, and control of their spectral positions is important to optimize the power available for the $f$-2$f$ technique. Post-fabrication trimming thus provides the option of being able to measure a device, determine its DW positions, and then trim its thickness to move them closer to the targeted values.

To demonstrate the possibility to post-process tune a DW through thickness trimming, we measure a DKS state obtained in a ring resonator with a starting thickness $H$=459~nm and $RW$=1040~nm and pumped at 976~nm in the fundamental transverse electric mode, while actively-cooled using a counter-propagative wave cross-polarized at 1060~nm, which allows for adiabiatic tuning onto the soliton state~\cite{zhouSolitonBurstsDeterministic2019,moilleUltraBroadbandSolitonMicrocomb2021}. Here, we proceed to monitor the DKS state as we trim the thickness by more than 30~nm, which corresponds to the full variation of Si$_3$N$_4$ thickness over a 100~mm \greg{wafer} (\cref{fig:1}). We use the same recipe as for the trimming calibration, namely a 20~sec RIE etch at 45~W RF power. By pumping the exact same mode, which is slightly shifted in wavelength for each trim \greg{step from approximately 307~THz (976~nm) to 309.5~THz (969~nm)}, a clear and constant red shift of the DW is observed for each trimming step. Assuming the previously calibrated trimming amount of 4~nm per step, the dispersion at each trimming step can be found theoretically, and the linear approximation of the DW position based on the zero-crossing of the integrated dispersion (\cref{fig:5}(b)-(c)) provides a reasonable estimate of the observed DW shift. \greg{It is important to note that for an asymmetric integrated dispersion profile (odd dispersion coefficients dominating), the linear approximation of predicting the DW position based on the zero-crossing of the integrated dispersion will present a discrepancy with respect to the actual DW frequency~\cite{cherenkov_dissipative_2017,jangObservationDispersiveWave2014}. To address this, we also present rigorous simulation data of the expected comb envelope} using the Lugiato Lefever equation model~\cite{moille_pyLLE}, where the only modified parameter at each trimming step is the \greg{dispersion profile, chosen to be consistent with the expected thickness, and with the same in-waveguide pump power and pump detuning of 180~mW and -1.932~GHz respectively}. The simulated comb spectra closely match the measured ones (Fig.~\ref{fig:5}(d)), confirming that the RIE etching process is mostly physical and does not affect the ring width. The power discrepancy between the theoretical and experimental DWs are due to the straight waveguide coupling system used for injecting and extracting light from the microring resonators, which exhibits a strong chromatic dependence and undercouples the resonator at short wavelengths~\cite{moille_broadband_2019}.

%   ____                 _           _             
%  / ___|___  _ __   ___| |_   _ ___(_) ___  _ __  
% | |   / _ \| '_ \ / __| | | | / __| |/ _ \| '_ \ 
% | |__| (_) | | | | (__| | |_| \__ \ | (_) | | | |
%  \____\___/|_| |_|\___|_|\__,_|___/_|\___/|_| |_|

In conclusion, we have demonstrated a dry etch technique that allows for fine control of the thickness of Si\textsubscript{3}N\textsubscript{4} microring resonators with a thickness step as low as 4~nm and over a range of 30~nm. The post-processing trimming does not significantly impact the Si$_3$N$_4$ surface roughness or resonator quality factors. By using an air-clad structure with oxide cladding still present at the chip facets, we are able to measure the resonator performance in-between trimming steps while retaining low insertion losses to lensed optical fibers. We demonstrate the utility of such fine trimming in microcomb engineering by showing how a dissipative Kerr soliton comb pumped in the 980~nm \greg{band} can have its dispersive wave tuned across a total range of 40~THz in few THz steps. This trimming approach has numerous potential uses in the development of microresonator frequency combs and related Kerr nonlinear optical devices. For example, it can enable multi-project wafer runs in which different thickness are needed across the wafer; recent demonstrations of four-wave mixing Bragg scattering~\cite{li_efficient_2016} and $\chi^{(3)}$ optical parametric oscillation~\cite{lu_-chip_2020} were realized for Si$_3$N$_4$ thicknesses (480~nm and 500~nm, respectively) that could be combined with the microcomb devices shown in this work. Finally, the trimming technique also makes it possible to compensate for the natural thickness variation of the Si$_3$N$_4$ material across a full wafer, so that the number of devices exhibiting a desired dispersion can be greatly increased.\\

See the supplementary material for information related to the insertion losses of the dual air/oxide cladding structure.

\noindent {\textsc{Acknowledgments}} - The authors acknowledge funding from the DARPA APHI and NIST-on-a-chip programs, and thank David Carlson and Feng Zhou for helpful comments. 

\noindent {\textsc{Data Availability}} - The data that support the findings of this study are available from the corresponding author upon reasonable request.

\bibliography{Moille_TrimmingComb}

\begin{thebibliography}{31}
\bibitem [{NIS()}]{NIST_disclaimer}%
  \BibitemOpen
  \href@noop {} {}\bibinfo {note} {Certain commercial products or names are
  identified to foster understanding. Such identification does not constitute
  recommendation or endorsement by the National Institute of Standards and
  Technology, nor is it intended to imply that the products or names identified
  are necessarily the best available for the purpose.}\BibitemShut {Stop}%
\end{thebibliography}
\clearpage
\section*{Supplementary Materials}
\beginsupplement
\subsection{Insertion Losses in Selectively Clad Devices -- Experimental Data}
\label{sub:IL}
\begin{figure}[!h]
    \includegraphics{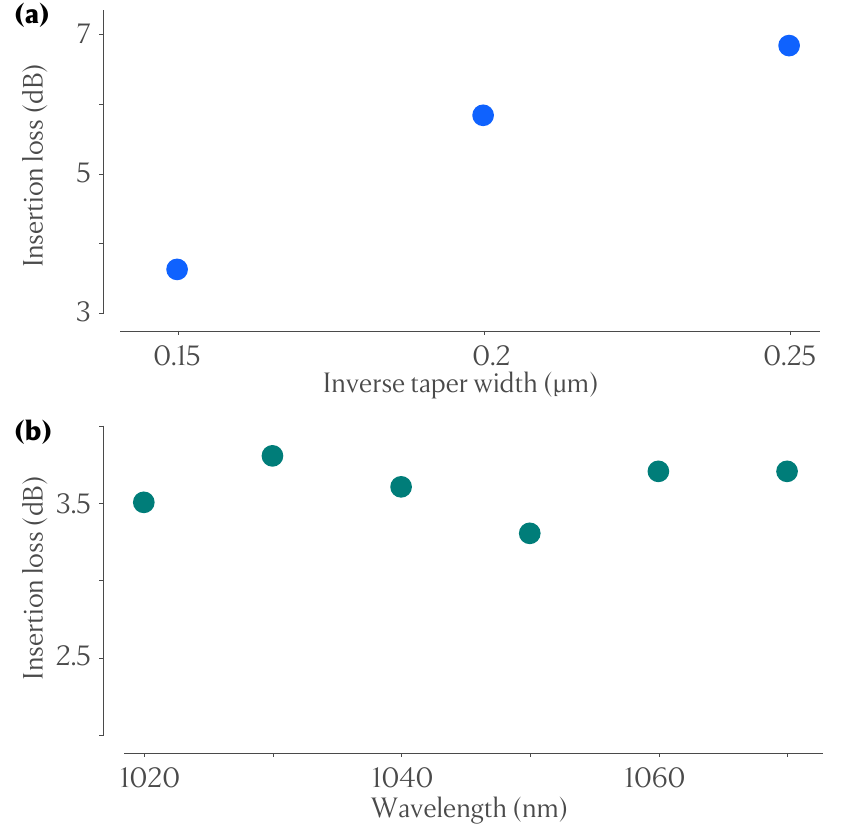}
    \caption{\label{supfig:1} \textbf{(a)} \ks{Experimentally measured} per facet insertion losses for the same material thickness as in \ks{Fig.~2(a) from the main text} and inverse taper widths from 150~nm to 250~nm. \textbf{(b)} Chromatic dispersion of the per facet insertion losses in the 1020~nm to 1070~nm band. The one standard deviation uncertainty in per facet insertion loss is at the $\pm0.5$~dB level and is due to measured variations in coupling for nominally identical devices.} 
\end{figure}

Our system utilizes a selective SiO$_2$ cladding, with the absence of cladding over the microresonator region allowing direct access for post-processing, while keeping an SiO$_2$ cladding at the facets that is useful for reduced insertion losses. Here we present experimental data supporting the simulations presented in the main article, which show that the insertion losses are lower than that in an air-clad system. We perform measurement of the insertion losses with the inverse taper width (\cref{supfig:1}(a)). As expected from the simulation, the insertion losses for the first order transverse electric mode increase with the inverse taper width, where the optimum value is at 150~nm width (smaller tapers are limited by the fabrication process; in particular, the ability to realize high-aspect ratios for thick Si$_3$N$_4$ films). Finally, we measured that such inverse tapers present very low chromatic dispersion in their losses across the pump band, and present similar insertion losses ($\approx$~3.5~dB per facet) across the range between 1020~nm to 1070~nm (\cref{supfig:1}(b)).\\

\subsection{Losses from the Cladding Transition}
\label{sub:ILclad}
\begin{figure}[t]
    \includegraphics[width=0.95\columnwidth]{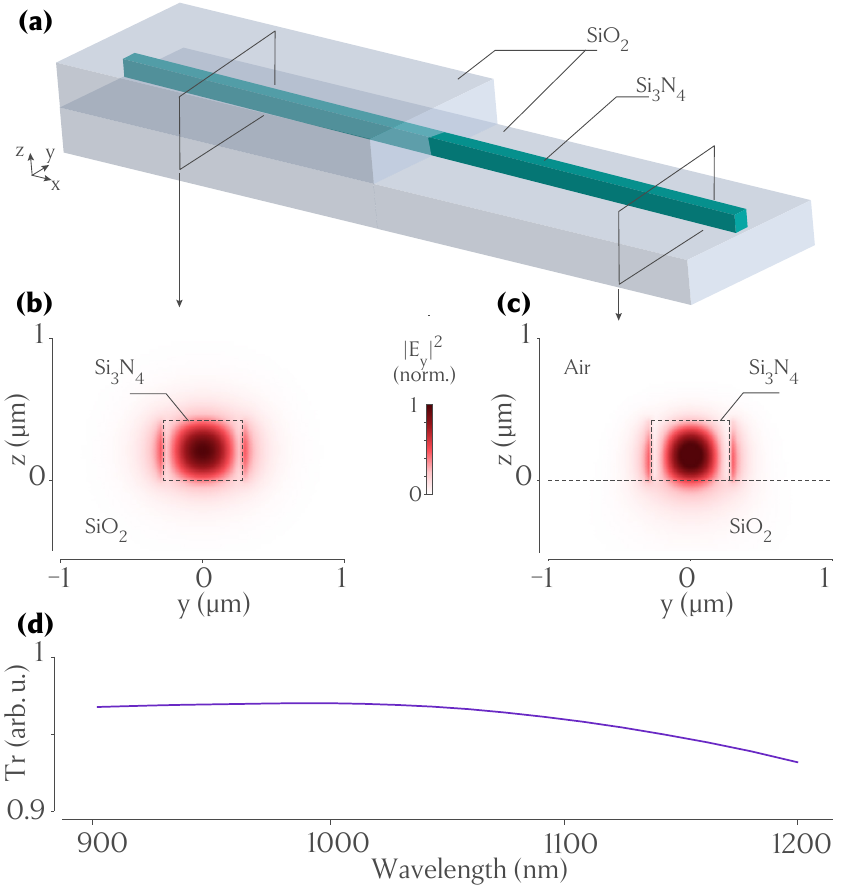}
    \caption{\label{supfig:2} \textbf{(a)} Schematic of silicon nitride waveguide cladding transition between oxide-clad and air-clad regions. \textbf{(b)} Mode profile of $y$-oriented electric field component of the first order transverse electric mode in the oxide clad region for a waveguide cross section of 550~nm width and 420~nm thickness. \textbf{(c)} Mode profile for the same polarization and waveguide cross-section, but with an air top/side cladding. \textbf{(d)} Finite-difference time-domain simulation results for the transmission between the oxide and air-clad regions.}
\end{figure}

In order to estimate the potential Frensel losses from the oxide-clad to air-clad transition (\cref{supfig:2}(a)), we perform finite-difference time-domain simulation of the structure, where a silicon nitride waveguide of 550~nm width and 420~nm thickness is halfway embedded in SiO\textsubscript{2} at the left edge, and then transitions to a waveguide that has only a bottom SiO$_2$ cladding and a top and side air cladding. Although the mode profile of the first order transverse electric mode presents some disparity between the two regions (\cref{supfig:2}(b)), due to the difference of refractive index gradient in the two regions that modifies the confinement of the mode, the simulated transmission from the oxide-clad region to the air-clad region remains well above 90~\% (\textit{i.e.} losses lower than 0.1~dB). This level of losses is negligible in comparison to the facet coupling loss.

\end{document}